\documentclass[a4paper]{aa}
\usepackage{natbib}
\usepackage{times}
\usepackage{epsfig}

\begin{document} 

\newcommand{\refcom}{\textbf}
\newcommand{\hd}{HD~81809}   
\newcommand{\lx}{$L_{\rm X}$}  
\newcommand{\es}{erg s$^{-1}$}   
\newcommand{\ecms}{erg~cm$^{-2}$~s$^{-1}$}
\newcommand{\halpha}{H$\alpha$}  
\newcommand{\hbeta}{H$\beta$}
\newcommand{\kms}{km~s$^{-1}$}   
\newcommand{\cmthree}{cm$^{-3}$}
\newcommand{\msun}{M$_{\odot}$} 
\newcommand{\xmm}{XMM-\emph{Newton}} 
\newcommand{\nh}{\mbox{$N({\rm H})$}}

\title{High-amplitude, long-term X-ray variability in the solar-type star
  HD~81809: the beginning of an  X-ray activity cycle?}

\author{F.\ Favata\inst{1} \and G.\ Micela\inst{2} \and S.\,L.
  Baliunas\inst{3} \and J.\,H.\,M.\,M. Schmitt\inst{4} \and M.
  G\"udel\inst{5} \and F.\,R. Harnden, Jr.\inst{6} \and\\ S.\
  Sciortino\inst{2} \and R.\,A. Stern\inst{7} } 

\institute{Astrophysics Division -- Research and Science Support
  Dept. of ESA, 
  Postbus 299, 2200 AG Noordwijk, The Netherlands
  \and 
  INAF -- Osservatorio Astronomico di Palermo, Piazza del Parlamento 1,
  I-90134 Palermo, Italy  
  \and
  Harvard-Smithsonian Center for Astrophysics, 60 Garden Street, MS
  15, Cambridge, MA 02138, USA
  \and
  Universit\"at Hamburg, Hamburger Sternwarte, Gojenbergsweg 112,
  21029 Hamburg, Germany
  \and
  Paul Scherrer Institut, W\"urenlingen and Villigen, CH-5235
  Switzerland
  \and
  Harvard-Smithsonian Center for Astrophysics, 60 Garden St.,
  Cambdridge 02138 MA, USA
  \and
  Lockheed Martin Advanced Technology Ctr., USA
  }
\offprints{F. Favata (Fabio.Favata@rssd.esa.int)}

\date{Received date / Accepted date}

\titlerunning{An X-ray cycle in HD~81809}
\authorrunning{F. Favata et~al.}

\abstract{ We present the initial results from our \xmm\ program aimed
  at searching for X-ray activity cycles in solar-type stars. \hd\ is
  a G2-type star (somewhat more evolved than the Sun, and with a less
  massive companion) with a pronounced 8.2 yr chromospheric cycle, as
  evident from from the Mt.\ Wilson program data. We present here the
  results from the initial 2.5 years of \xmm\ observations, showing
  that large amplitude (a factor of $\simeq 10$) modulation is present
  in the X-ray luminosity, with a clearly defined maximum in mid 2002
  and a steady decrease since then. The maximum of the chromospheric
  cycle took place in 2001; if the observed X-ray variability is the
  initial part of an X-ray cycle, this could imply a phase shift
  between chromospheric and coronal activity, although the current
  descent into chromospheric cycle minimum is well reflected into the
  star's X-ray luminosity. The observations presented here provide
  clear evidence for the presence of large amplitude X-ray variability
  coherent with the activity cycle in the chromosphere in a star other
  than the Sun.  \keywords{Stars: X-rays} }

\maketitle

\section{Introduction}
\label{sec:intro}

The 11 year cycle is perhaps the oldest known manifestation of the
Sun's magnetic activity, having first been noted by Schwabe in 1843 as
a periodic modulation of the number of sunspots. Subsequently, most
activity indicators have been observed to follow a similar cyclical
variation, with the amplitude of the modulation dependent on the
indicator used. On cool stars other than the Sun, the detection of
cycles had to wait for the foresight of O. Wilson, who started a
long-term monitoring of a Ca\,{\sc ii} H\&K activity indicator (the
``$S$ index'') in a significant number of stars, using the Mt. Wilson
100 inch telescope. An analysis of the vast amount of data from the
Mt. Wilson program, now covering nearly 40 years of observations
(\citealp{bds+95}), shows that solar-like cycles are present in many
stars, although some stars show no detected variability (perhaps being
in a ``Maunder-like'' state) while others show significant
non-periodic variability. In the Sun, the amplitude of the cyclical S
index modulation is a factor of about 2, while the amplitude in X-rays
is much stronger, i.e.\ a factor of 100 in the Yohkoh 0.73--2.5 keV
band.

Yet evidence of cyclical variability in X-rays in stars other than the
Sun has only become available very recently. In fact (with the recent
exception of 61~Cyg), the X-ray observations of the few stars for
which sufficient data exist suggest that their X-ray luminosity is
relatively stable over long-term intervals, as discussed e.g. by
\cite*{ste98b}. Observations of homogeneous samples of active stars,
e.g.\ of the Hyades, show little variation in \lx\ across the $\simeq
10$-year separation between the \emph{Einstein} and ROSAT PSPC
surveys, with 90\% of the stars showing less than a factor of 2
variability. In these stellar samples, however, the median activity
level is much higher than the Sun's, by 2 or more dex: stars at this
activity level typically do not show cycles in Ca\,{\sc ii}, instead
varying irregularly, so that perhaps the lack of X-ray modulation is
not surprising.  This result was however confirmed by \cite*{ls2004}
on a volume-limited sample of solar-type stars using ROSAT All-Sky
Survey (RASS) and pointed observations. Weak statistical evidence for
the presence of solar-like cycles in less active stars was derived by
\cite*{hss96} using the RASS observations of the stars in the Mount
Wilson program, looking at the deviations from a ``mean'' X-ray
luminosity for each star as a function of the Ca\,{\sc ii} cycle phase.

In their analysis of X-ray variability properties of solar mass stars
\cite*{mmp+2002} found that, in relatively quiet stars, amplitude
variations increase with time scales, and interpreted this as an
indication of the presence of solar-like cycles in stars with X-ray
activity of the same order of that of the Sun. The comparison between
the Sun and nearby stars is consistent with a fraction of moderately
active stars ($\overline{L}_{\rm X} < 10^{28}$ erg/sec) having X-ray
variability similar to the Sun, while more active stars lack
solar-like cyclic coronal activity (\citealp{mm2003}). 

More recently, \cite{hsb+2003} presented evidence of X-ray luminosity
modulation in 61 Cyg A and B (K5V and K7V) over 4.5 years, well
correlated with their $S$ index, strongly suggestive of an activity
cycle in X-rays. The observed modulation amplitude (using ROSAT HRI
data, which prevented a study of the spectral evolution) is a factor
of 2.5.

In this paper we present the results of the first 2.5 yr of the \hd\ 
\xmm\ monitoring program, showing for the first time in a star other
than the Sun strong evidence for long-term coherent X-ray variability
with an amplitude a factor of about 10, consistent with the cyclic
amplitude modulation in the Sun.

\section{Characteristics of \hd}

To search for the presence of clearly visible cycles in the X-ray
activity of solar type stars we have selected what appeared to be the
best target from the Mt. Wilson sample, \hd, which has a very clearly
defined cycle in Ca\,{\sc ii}, with a sufficiently short period
(8.2~yr, \citealp{bds+95}), and sufficient X-ray flux at Earth to
permit its efficient observation. We have started a long-term
monitoring program of this star with \xmm, observing it once every 6
months, with the aim, over the \xmm\ lifetime, of searching for
clearly defined cycles and compare their characteristics with the ones
of the Ca\,{\sc ii} cycle in the same star as well as with the
characteristics of the solar cycle.

\hd, was once considered a good ``solar analog'', but in fact it is a
visual binary system, with a maximum separation of 0.4 arcsec and a
period of about 35 yr (\citealp{pou2000}), so that tidal effects are
likely to be negligible. The masses of the two components are $M_1 =
1.7 \pm 0.64\, M_\odot$ and $M_2 = 1.0 \pm 0.25\, M_\odot$, with
spectral types G2 and G9 and apparent magnitudes $V_1 = 5.8$ and $V_2
= 6.8$ respectively. Both components are slow rotators, with $v\sin i
= 3$ km/s (\citealp{sod82}. The Hipparcos parallax appears not to be
correct because of the system's binary nature, and has been revised
(\citealp{sod99}) to $\pi = 29.1 \pm 1.1$ mas, which we adopt here.
Such parallax (together with the apparent magnitude) implies that the
stars are not on the main sequence, but rather subgiants.

\begin{figure}
  \begin{center} 
    \leavevmode 
        \epsfig{file=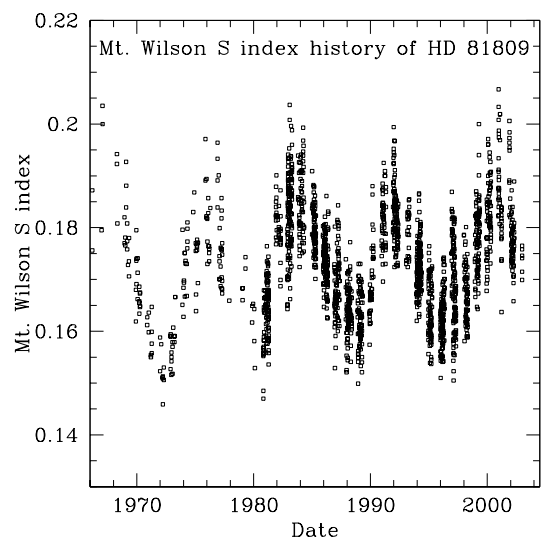, width=7.4cm}
  \caption{Evolution of the Mt. Wilson $S$ index of \hd\ from 1966 to
    the end of 2002 (see \citealp{bds+95} for details). }
  \label{fig:his}
  \end{center}
\end{figure}

In the Ca\,{\sc ii} $S$ index \hd\ shows a very clear cyclical
behavior, with a shape rather similar to that of the Sun and an 8.2
year period. All Mt. Wilson $S$ index observations are plotted in
Fig.~\ref{fig:his}. The binary is not resolved, either in the Mt.
Wilson data nor in the X-ray observations described below.  However,
the very clear and regular modulation evident in Fig.~\ref{fig:his}
points to the activity to be dominated by one of the two components.

\section{Observations}
\label{sec:obs}

Monitoring observations of \hd, executed every 6 months by \xmm,
started in April 2001. Each 'snapshot' is a nominal 7 ks exposure. All
observations up to May 2003 were taken with the medium filter, while
the last observation was taken with the thick filter (following a
policy change for the observatory).

All data sets were processed in the same way using the SAS V. 5.4.1
package. High background time intervals were removed prior to further
processing and spectra were extracted for each observation, for the pn
as well as for each of the MOS cameras. The pn and MOS spectra were
simultaneously fit, using {\sc xspec}, with a single temperature
optically thin plasma model ({\sc apec}). The metal abundance was
frozen at $Z = 0.3\,Z_\odot$, to reduce the number of free variable
and allow as consistent as possible a comparison among the different
data sets. An interstellar absorption component was not necessary to
obtain a good fit. For all of the observations, pn and MOS data
produced consistent spectral parameters.

\begin{figure}
  \begin{center} 
    \leavevmode 
        \epsfig{file=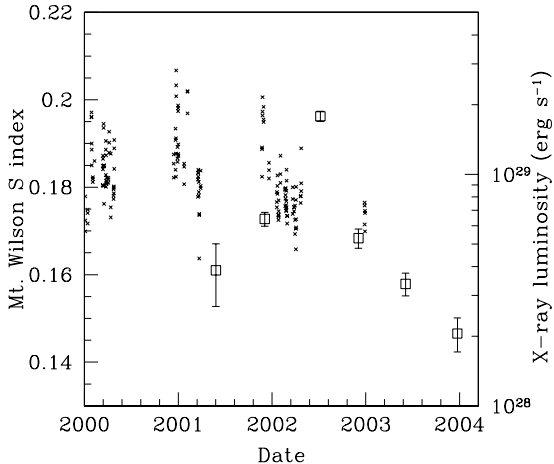, width=7.4cm, bbllx=14pt,
     bblly=154pt, bburx=585pt, bbury=640pt, clip=}
  \caption{Evolution of the mean (averaged over the duration of each
    \xmm\ observation, ca. 2 hr) X-ray luminosity in the 0.2--2.5 keV
    band (right hand scale) of \hd\ from April 2001 to November 2003
    together with the available Mt. Wilson $S$ index measurements
    (left hand scale). }
  \label{fig:sim}
  \end{center}
\end{figure}

In all cases except for the June 2002 observation the single
temperature model provided a good fit to the data, with temperatures
consistently around 0.35 keV.  Nevertheless, a single temperature fit
was also performed for the June 2002 data to have a homogeneous data
set. X-ray luminosities were computed for both a ROSAT-like 0.2--2.5
keV band and for a Yohkoh-like 0.73--2.5 band. The resulting spectral
parameters are reported in Table~\ref{tab:par}, and a long-term light
curve of the X-ray luminosity of \hd\ is plotted (together with the
$S$ index) in Fig.~\ref{fig:sim}.

\begin{table*}[!thbp]
  \begin{center}
    \caption{Best-fit spectral parameters for the 6 \xmm\ observations
      of \hd\ discussed here. Count rates in cts s$^{-1}$,
        luminosity in erg s$^{-1}$, temperature of the plasma in
        keV.}  \leavevmode
    \begin{tabular}{ccccc}
Date & pn rate & \lx\ (0.2--2.5 keV) & \lx\ (0.73--3.5 keV) & $kT$ \\\hline
2001-04-25 & $0.1830 \pm 0.055$ & $3.85 \times 10^{28}$ & $1.46 \times 10^{28}$& 0.34 \\
2001-11-01 & $0.3418 \pm 0.024$ & $6.42 \times 10^{28}$ & $2.91 \times 10^{28}$& 0.40 \\
2002-06-06 & $0.7456 \pm 0.037$ & $1.78 \times 10^{29}$ & $8.88 \times 10^{29}$& 0.81 \\
2002-11-02 & $0.2455 \pm 0.024$ & $5.31 \times 10^{28}$ & $2.31 \times 10^{28}$& 0.39 \\
2003-05-03 & $0.1522 \pm 0.017$ & $3.37 \times 10^{28}$ & $1.36 \times 10^{28}$& 0.36 \\
2003-11-22 & $0.0744 \pm 0.012$ & $2.05 \times 10^{28}$ & $7.33 \times 10^{27}$& 0.33 \\
    \end{tabular}
    \label{tab:par}
  \end{center}
\end{table*}

\begin{figure}
  \begin{center} 
    \leavevmode 
        \epsfig{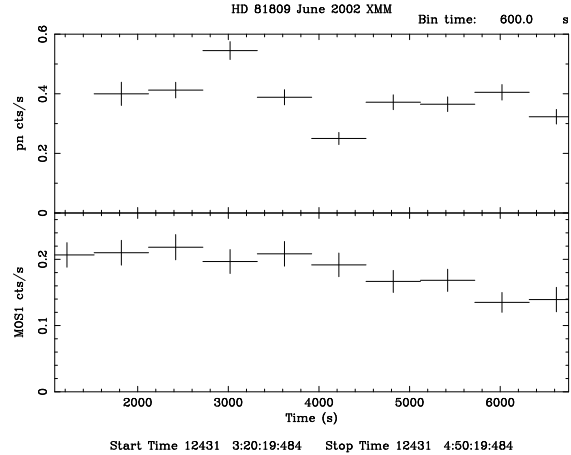}
  \caption{Light curves (at 600 s bins) of \hd\ during the June 2002
    \xmm\ observation. Top panel, pn data, bottom panel, MOS2 data. }
  \label{fig:lc}
  \end{center}
\end{figure}

During the June 2002 observation the X-ray luminosity of \hd\ is at
its maximum, and to obtain a good fit to the spectrum a second
temperature component, as well as a different coronal abundance, were
needed. The best fit temperatures were $T_1 = 0.51$ keV, $T_2 = 1.3$
keV, i.e. even the lower temperature component is significantly hotter
than during the other observations. The relative emission measure of
the two components was comparable, with $E\!M_1/E\!M_2 = 0.81$, and a
best-fit coronal abundance $Z \simeq 0.2\,Z_\odot$. The resulting
X-ray luminosity is close (within $\simeq 10$\,\%) to the one
determined through the single temperature fit, justifying the approach
used for a homogeneous comparison. Clearly, the ``coronal state'' of
\hd\ during this observation was different than during the other
observations, with a significantly harder spectrum. The short duration
of the June 2002 observation makes it difficult to determine if \hd\ 
was undergoing a flaring event.  The June 2002 observation
(Fig.~\ref{fig:lc}) shows significant variability, with the count rate
decreasing rather irregularly during the observation, but without a
clear decay (or rise) which could be associated with a ``classic''
flare. Also, the hardness ratio does not change significantly during
the observation, pointing to a constant (rather than decreasing, as
expected in the decay phase of a flare) plasma temperature.

Most solar flares have typical duration of tens of minutes (although
some very long ones, up to a day, have been observed), and such short
event would have a very clear signature in the two hour span of the
observation.  Long-duration flares, up to days, have been observed in
very active stars; while \hd's activity level is higher than solar,
even when considering surface X-ray flux rather than luminosity, it
still is much less active than the very active stars in which long
duration flaring events have so far been observed to be common.

\section{Discussion}

Very significant X-ray long term variability is present in \hd\ 
(Fig.~\ref{fig:sim}). This alone is a significant result, as in all
X-ray active stars monitored with a sufficient time base the long term
variability is very small (typically a factor $\le 2$, see
\citealp{ste98b}). The X-ray luminosity of \hd, which is not a
high-activity star, varies by more than an order of magnitude in two
years, with a systematic pattern. If the June 2002 observation were
discounted as due to a long-duration flare, the total light curve
amplitude would still be a factor of approximately 5. The amplitude of
cycle-related modulation in the X-ray luminosity is thus higher in
\hd\ than that recently reported for 61~Cyg (a factor of 2.5,
\citealp{hsb+2003}).

\begin{figure}
  \begin{center} 
    \leavevmode 
        \epsfig{file=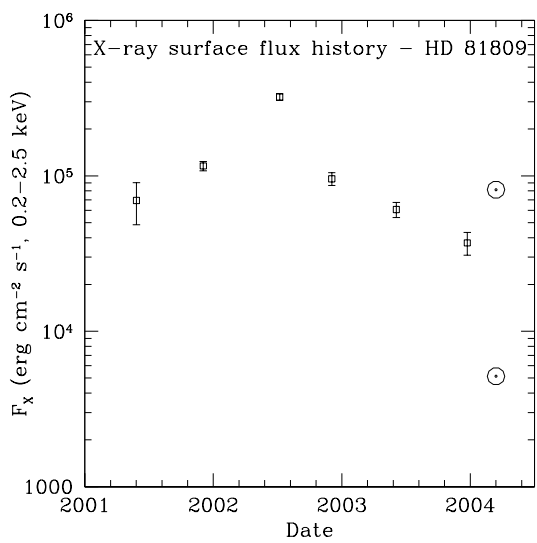, width=7.4cm}
  \caption{Evolution of the X-ray surface flux (in the 0.2--2.5 keV
    band) of \hd\ from April 2001 to November 2003. At the right of
    the plot also the typical X-ray surface flux of the Sun at minimum
    and maximum of the cycle are plotted. }
  \label{fig:lxev}
  \end{center}
\end{figure}

While more luminous than the Sun in absolute terms, \hd\ is a
subgiant, with a typical radius of $R \simeq 3 R_\odot$. The observed
X-ray luminosity range corresponds(considering only the primary star)
to a surface flux $4.6 \ge \log F_{\rm X} \ge 5.5$, somewhat higher
than the corresponding solar values. At the same time, the Ca\,{\sc
  ii} surface flux $R_{\rm HK}$ is almost identical in \hd\ and in the
Sun ($\log R_{\rm HK} = -4.90$ versus $\log R_{\rm HK} = -4.92$
respectively, \citealp{bds+95}). The X-ray surface flux of \hd\ is
plotted, in Fig.~\ref{fig:lxev}, together with the range of values
observed in the Sun during the cycle. The coronal temperature of \hd\ 
is (except for the June 2002 observation) rather constant, and
somewhat higher than the coronal temperature of the disk-integrated
Sun. The coronal emission measure of the Sun (when filtered through
the response of a CCD X-ray detector, see \citealp{por+99} for
details) has a bulk component whose temperature varies little, between
0.16 and 0.19 keV between the minimum and the maximum of the cycle,
with the addition of a hotter component ($T = 0.49$ keV), which is
only present at cycle maximum.  Whether the June 2002 is part of the
cycle behavior of \hd\ or is an exceptional event is something which
can only be clarified by further observations.

Figure~\ref{fig:sim} shows our determination of the X-ray luminosity
of \hd\ plotted together with the last years of $S$ index measurements
(due to wildfires that threatened the Mt. Wilson area, very few
observations were obtained in 2003).  The Ca\,{\sc ii} chromospheric
cycle had a well defined maximum in 2001, and it is now in its
descending phase; based on the previous cycles this will likely last
well into 2005. On the other hand the X-ray luminosity suggest a cycle
maximum corresponding to the June 2002 observation (independent from
whether or not a flare occurred).  Taken at face value, this would
seem to imply a phase shift between the chromospheric and the coronal
activity cycles. The last three determinations of the X-ray luminosity
are consistently decreasing, as expected on the basis of the
chromospheric cycle of the star. The lack of $S$ index values in 2003
prevents us from performing a correlation analysis like that of
\cite*{hsb+2003}.

\section{Conclusions}
\label{sec:concl}

We have presented clear evidence of long-term variations of the X-ray
luminosity in \hd, a solar-type star with a well defined cycle in its
chromospheric activity. The variations thus far determined have an
amplitude of a factor of 10 (over the 2.5 years covered by the
observations), comparable to the variations seen in the X-ray
luminosity of the Sun during the solar cycle. These variations suggest
the beginning of a cycle; while the observed X-ray maximum appears
somewhat offset (by about 1 year) from the chromospheric one, the
current descending phase of the chromospheric cycle is well reflected
in \hd's decreasing X-ray luminosity over the last two years.

\hd\ is the subject of a long-term monitoring program performed with
the \xmm\ observatory (which also includes $\alpha$ Cen and 61~Cyg),
of which we are presenting here the first (still perforce preliminary)
results. A further two years of observations (at six months cadence)
are already planned on the same target, and we will re-propose the
target for every AO, to ensure the continuous monitoring. While the
nature of this program is such that a few more years will be necessary
before a detailed analysis can be performed, the initial results
presented here show that our initial choice of targets was a good one;
we now can state with some confidence that the Sun is not the only
solar-type star for which large amplitude X-ray variability coherent
with the activity cycle in the chromosphere is present.  The
continuation of our monitoring program on \hd\ will allow to compare
the characteristics of its coronal cycles with the solar one.

\begin{acknowledgements}
  
  GM, SS acknowledge the partial support of ASI, MIUR. This paper is
  based on observations obtained with \xmm, an ESA science mission
  with instruments and contributions directly funded by ESA Member
  States and the USA (NASA). Partial support of the observations at
  Mt.  Wilson is acknowledged from the Air Force Office of Scientific
  Research, AF 49620-02-1-0194. We would like to thank R.\,A. Donahue
  as well as the colleagues who have supported the initial \xmm\ 
  proposal, namely A.  Collier Cameron, R. Rosner, K.  Strassmeier, F.
  Walter.

\end{acknowledgements}


\end{document}